# Magnetic domain structure and dynamics in interacting ferromagnetic stacks with perpendicular anisotropy

S. Wiebel, J.-P. Jamet, N. Vernier, A. Mougin, J. Ferré

*Laboratoire de Physique des Solides, UMR CNRS 8502, Université Paris-Sud, 91405 Orsay, France*

V. Baltz, B. Rodmacq, B. Dieny

*SPINTEC, URA CNRS/CEA 2512, CEA-Grenoble Cedex 9, France*

The time and field dependence of the magnetic domain structure at magnetization reversal were investigated by Kerr microscopy in interacting ferromagnetic Co/Pt multilayers with perpendicular anisotropy. Large local inhomogeneous magnetostatic fields favor mirroring domain structures and domain decoration by rings of opposite magnetization. The long range nature of these magnetostatic interactions gives rise to ultra-slow dynamics even in zero applied field, i.e. it affects the long time domain stability. Due to this additionnal interaction field, the magnetization reversal under short magnetic field pulses differs markedly from the well-known slow dynamic behavior. Namely, in high field, the magnetization of the coupled harder layer has been observed to reverse more rapidly by domain wall motion than the softer layer alone.

# 1. Introduction

Up to recently, magnetic recording media or spintronic devices are essentially based on metallic thin film structures with in-plane magnetic anisotropy. Drastic limitations appear in associated patterned media when the dimension of magnetic elements goes down to nano-sizes. Perfect two states systems with perpendicular anisotropy are thus attractive since they are expected to allow an increase of the media storage density [1], and improve the stability and the reliability of magnetic tunnel junction elements [2]. There is now a growing up interest for investigating such ultrathin film structures with perpendicular magnetic anisotropy that may be used in multilevel magnetic recording [3] or in exchange bias structures for designing magnetic random access memories or magnetic read heads [3-6]. Multilevel perpendicular recording media structure can be realized by stacking ultrathin magnetic multilayers and separating them by non-magnetic spacer layers; each magnetic multilayer then exhibits perfect square hysteresis loops with adjustable coercivity. The magnetic properties of ultrathin multilayers can be better adjusted than in single layers; in particular, thanks to prevailing interface contributions, the perpendicular anisotropy can be preserved up to relatively thick multilayers. Some magnetic memories are built with at least one hard and one soft ferromagnetic layer separated by a non-magnetic spacer layer. The spacer layer is either metallic in giant magnetoresistance (GMR) or insulating in tunnel magnetoresistance (TMR) magnetoelectronic devices.

In all cases, the knowledge of static and dynamic behavior that outcomes from the coupling between magnetic layers is mandatory to improve the stability and reliability in real systems. Only few results have been reported so far on dynamics of the magnetization reversal except for coupled in-plane magnetized layers [7]. From a general point of view, two situations have to be considered, both for systems with in-plane or out-of plane anisotropy: the interaction between two magnetically saturated layers, and the far more complex case of an inhomogeneous interaction between domain structures present in each layer.

It is known that either in-plane or out-of-plane magnetically saturated layers separated by a pinhole free non-magnetic metallic spacer layer are coupled by RKKY oscillatory interlayer exchange interaction that decreases when increasing the spacer thickness [8-10]. Moreover, a conformal magnetostatic interlayer coupling, due to the so-called orange peel (op) or Néel effect, can be induced by correlated roughness of ferromagnetic-spacer interfaces [11-13]; this interaction also vanishes progressively when increasing the spacer layer thickness. The sum of the magnetostatic and exchange interactions induces a bias shift of the hysteresis loop, which is relatively small for a spacer layer thickness of a few nanometers [13]. Depending of its sign with respect to the initial saturation direction, the bias field, $H_b$, increases or decreases the nucleation field value of the soft layer.

Magnetostatically-induced stray fields become more efficient in non-uniformly magnetized films, i.e. during the magnetization reversal process or in non-fully saturated structures. Their study is thus a major issue for basic physics as well as for applications, as already shown in GMR or TMR in-plane magnetized structures. For example, this coupling allows explaining why pseudo-spin valves or double coercivity tunnel junction's memories may be erased progressively since the hard layer demagnetizes with repeated reversal of the soft layer by field cycling [14]. For in-plane magnetized structures, it has been shown that this relaxation phenomenon comes from the replication of the soft layer domain walls inside the hard layer and vice-versa [15]. In other words, even under moderate fields, mirrored domains can be stabilized in the hard and soft layers with in plane anisotropy through the mutual imprint of domain walls [16]. In that case, the stray field created by Néel walls in one layer lowers the nucleation field in the other one. Thus, some questions obviously arise for ultrathin film coupled structures with perpendicular anisotropy : (i) Have we such domain replication when cycling the field, since domain walls might not be of the same type ? (ii) Is a created domain structure stable enough at long time to ascertain reliability ? This requires to consider slow « creep » dynamics [17, 18] in the data analysis. (iii) Is the role of interactions the same at all time scales ? To our knowledge, these two last points were not investigated so far even in systems with in-plane anisotropy.

The aim of this article is to treat magnetostatically-induced phenomena during magnetization reversal in a structure consisting of a hard and a soft ferromagnetic Co/Pt multilayer stack with perpendicular anisotropy, separated by a thicker non-magnetic Pt spacer layer. As for systems with in-plane magnetic anisotropy, the room temperature magnetic domain structure in the two stacks was visualized by magneto-optical microscopy [20]. This method allows investigating time dependent phenomena, related to domain nucleation and wall propagation [18]. We have previously reported on a magnetostatically-induced effect, the decoration of a hard magnetic domain at remanence by a ring of opposite magnetization in the soft stack [19]. In the present article, the dynamics of expansion or shrinking of this ring under the application of a weak field will be investigated and interpreted in the frame of the domain wall motion creep theory. Moreover, we will focus on domain mirroring between layers. The associated slow dynamics of the soft stack and fast response of both stacks will be studied to give some insight on the long-term stability and writing process in memories, respectively.

## 2. Sample properties

*2.1. Preparation and general magnetic properties of the bi-stack structure*

We studied the $S_H$/Pt(4 nm)/$S_S$/SiO$_2$/Si film structure, consisting of a hard $S_H$: [Pt(1.8 nm)/Co(0.6 nm)]$_4$ and a soft $S_S$: [Co(0.6 nm)/Pt(1.8 nm)]$_2$ ferromagnetic multilayer stack, separated by a Pt(4 nm) spacer layer, and deposited on a thermally oxidized silicon substrate. The sample was prepared by dc magnetron sputtering at room temperature under 2.5. 10$^{-3}$ mbar Ar pressure, with rates of about 0.05 nm/s [19]. The base pressure was around 4. 10$^{-8}$ mbar.

In this section, we report on the magnetism of the $S_H$/Pt(4 nm)/$S_S$ bi-stack sample. The Pt(1.8 nm) spacer within a stack is thin enough to preserve a large enough ferromagnetic exchange coupling ($H_{ex} \approx$ 100 Oe) that promotes a magnetic cooperative behavior and perpendicular anisotropy within $S_H$ and $S_S$ [21]. In counterpart, the rather large inter-stack Pt(4 nm) spacer layer thickness provides only a negligable RKKY interaction between $S_H$ and $S_S$ [13]. The relatively high positive value of the anisotropy constant K = K$_V$ + K$_S$, respectively equal to 1.71 x 10$^6$ and 1.29 x 10$^6$ erg/cm$^2$ for $S_H$ and $S_S$, gives rise to a net perpendicular magnetic anisotropy. As expected, polar Kerr rotation hysteresis loop in perpendicular applied field (Fig. 1) shows two successive jumps at $H_C^S$ = 106 Oe, and $H_C^H$ = 320 Oe, respectively for the soft and hard layer [19]. These values are rather close to the coercive fields we deduced in the corresponding single stack samples, $S_S$/SiO$_2$/Si and $S_H$ /Pt(4 nm)/SiO$_2$/Si, when measured in the same dynamic conditions (field sweeping rate = 500 Oe/s). Magnetization reversal dynamics in the bi-stack are evidenced by an increase of both $H_C^S$ and $H_C^H$ of 30 Oe per decade for a change in field sweeping rate between 0.5 and 50 kOe/s. As previously mentioned [19], after saturating the sample in a positive field and describing a minor hysteresis loop (Fig. 1), the negative coercive field value is found larger than the positive one. The corresponding bias field, $H_b$ = 8 Oe, is assigned to an orange peel magnetostatic interaction that favors a parallel alignment of the magnetization in both stacks and delays nucleation when decreasing the field from a positive saturated state.

As reported in our previous letter [19], an up-up and down-down ribbon-like domain structure is observed in the ac-demagnetized state of the sample. The absence of intermediate (down-up or up-down) magnetization states proves the existence of large local inter-stack magnetostatic interactions that induce a perfect ferromagnetic matching of domains and walls in $S_H$ and $S_S$.

### 2.2. Domain structure and wall dynamics in $S_S$.

The field-induced domain patterns have been imaged by high-resolution polar magneto-optical Kerr effect (PMOKE) microscopy (resolution: 0.4 μm) using a green LED light source. In our experiments, the sample was first magnetized in a positive field $H$, and the reference magneto-optical image in the up-saturated state recorded. The sample was subsequently submitted to a fixed negative field - $H$. To investigate dynamics, two distinct procedures have been used for recording PMOKE images of field-induced states:

- For fast most of the following experiments, and when nothing is indicated, we used the procedure (i) : the field was switched off to zero to observe the frozen remanent state associated to the field - $H$, applied during a certain time t. Then, PMOKE snapshots of the frozen remanent final state are recorded for increasing times t.
- In the case of ultra-slow dynamics, when the wall position does not move much during the image acquisition time, the (ii) procedure was sometimes used, where PMOKE snapshots of the domain pattern were recorded during the application of a field - $H$ for different times t. In order to improve the quality and contrast of the magnetic patterns, the images of the reference magnetically saturated state and of the final remanent or in-field states are accumulated on 16 bits and subtracted from each other by image digital processing.

As in the single stack $S_S$ sample, the field-induced magnetization reversal of $S_S$ from the saturated up-up spin state to the down-up state in the bi-stack structure (in the following down-up or up-down states will correspond to $S_S$ in the down or up-magnetized state and $S_H$ in the up or down-magnetized state, respectively) occurs by quite-uniform domain wall propagation initiated at only few nucleation sites (Fig. 2) [18]. As found for the single stack $S_S$ sample, after nucleation, the wall moves quasi-instantaneously (the lag time is estimated here to be smaller than 0.1 µs) as soon as the field is applied. At low field, the wall roughness is always weak, allowing to describe its motion in the frame of the « creep » theory, generally valid for the motion of a quasi 1D interface in a weakly disordered medium [17, 18]. The wall is then pinned cooperatively by a large assembly of weak defects. In that case, the low field dependence of the mean wall velocity, v, is expressed by

$$v = v_0 \exp[- A \, (H_{crit} / H)^{1/4}] \tag{1}$$

where $H_{crit}$ stands for the critical depinning field, and A depends of the degree of disorder. At very low field, the domain wall moves by rare and sudden events involving rather large Barkhausen volumes [22]; thus, only the mean wall velocity, over extended length and time scales, has a physical meaning. In the bi-stack structure, the v(H) variation of the wall in $S_S$ is shown on Fig. 3. The depinning field, $H_{crit}$, that separates the thermally activated regime from the viscous regime is estimated here to be 210 Oe. At low field, in the "creep" regime, the typical mean domain wall velocity is 1.5 µm/s for $H$ = 39 Oe and 2.2 x $10^{-5}$ µm/s for $H$ = 16 Oe. The plot of v versus $H^{-1/4}$ at low field (Fig. 4) fits well the « creep » expression (1). One deduces $A(H_{crit})^{1/4}$ = -103 and $Ln(v_0)$ = 30.2. One generally expects that the « creep » law fails at large field [17], i.e. here above 160 Oe, a value that obviously remains still below $H_{crit}$ = 210 Oe (Fig. 4).

As also found in the single stack $S_S$ sample, and more generally in other ultrathin films with perpendicular anisotropy, the final sequence of the slow magnetization reversal in $S_S$ (H < 52 Oe) brings into play regions with non-reversed up-magnetized channels, also called vestigial 360° winding walls (Fig. 5) [23]. They are built by two distinct mechanisms. In the first one, a few domains are nucleated far away from each other and after expansion, their walls come close together, forming a connected array of channels that is stabilized by intralayer magnetostatic forces (Fig. 5a) [24]. For the second mechanism, as soon as a wall is touching an extrinsic defect during its motion, it coils around it to finally form two facing walls preferentially oriented along the mean displacement vector. This leaves up-magnetized channels that are pinned at one of their extremity (Fig. 5a). For the first type, the facing walls connected structure remain stable over a long time since it is stabilized by a repulsive interaction. Second type up-magnetized channels can shrink after a time delay through their sudden thermally activated depinning on the defect located at its extremity (Fig. 5b). Since the magnetostatic interaction is weak for such ultrathin magnetic stack, a small increase of the field value is sufficient to wash out the second type of up-magnetized channels (Fig. 5c) leaving only a short filamentary up-domain (Fig. 5d). Again, this small domain will disappear later by depinning at one of its extremity. Thus, at low field sweeping rate, and as globally shown on the hysteresis loop (Fig. 1), these vestigial domains in $S_S$, will have absolutely no incidence on the initiation of the magnetization reversal in $S_H$. Starting from the situation depicted on Fig. 5d and reversing the field to a positive value, one observes that the small subsisting non-reversed channel (Fig. 5d) acts now as a nucleated area from which an up-magnetized domain (Fig. 5e) starts to expand by wall motion. Repeating this procedure, but starting now from a down-saturated $S_S$ state prepared under a larger field of - 100 Oe, we evidenced that the nucleation of $S_S$ up-magnetized domains in positive field occurs at places that are non-correlated with vestigial 360° walls (Fig. 5a-d).

### 3. Dipolar coupling and domain decoration in the remanent state

After saturating the sample magnetization in the up-up state, a short negative field pulse was applied to entirely reverse the magnetization downwards in $S_S$ and to initiate the reversal of small down-down magnetized domain in $S_H$. Just after switching off the field to zero, the down-down (black) domain is unchanged, but an up-up (white) magnetized decoration ring is found to expand rapidly (in a time shorter than the image acquisition time of some seconds) around it over a short distance (~ 1 µm) (Fig. 6). Non-reversed matched up-up (white) magnetized channels are also maintained inside the central down-down (black) domain. These narrow channels, limited laterally by facing walls, are stabilized by the intra-stack dipolar interaction [23, 24], but as demonstrated later, much more by stronger local inter-stack magnetostatic stray fields. Their width (Fig. 6) is overestimated because of the limited optical resolution of our set-up; their physical width is smaller than 0.4 µm. The jaggedness of the domain wall in $S_H$ is also duplicated instantaneously on the $S_S$ wall by the strong inter-stack magnetostatic interaction.

The most intriguing feature here is surely the appearance of this up-up magnetized (white) ring surrounding or "decorating" the central down-down (black) domain after switching off the field to zero. Due to a weaker pinning in the soft stack $S_S$, the external wall of the ring becomes smoother than the internal coupled one. Apart from the inside of the decoration ring, the $S_S$ magnetization points down, providing a down-up (gray) state.

It has to be emphasized that the creation of a regular decoration ring needs sometimes a special preparation procedure. As discussed later, after switching off the field, the up-up (white) ring state is initiated first in channels where local magnetostatic fields are more efficient (Fig. 6). The picture of a regular expansion of the up-up state from channels to form a domain ring is not always fulfilled at short experimental time scale. The propagation of the wall, initially formed inside a channel, can be pinned at points located just around the black domain. Such a situation is depicted in Fig. 7, where the white ring is only present over a limited portion of the black domain perimeter. In order to surely generate a uniform ring around the black domain in the remanent state, one applies first a small positive field (~ 20 Oe) to expand the ring, then a negative field ($H$ ~ - 70 Oe) to shrink it, and finally switch off the field.

A simple sketch of the formation of the decorated ring at remanence, excluding the channel formation, is depicted in Fig. 8. Let us start from the bi-stack domain state in a large negative field (Fig. 8a); $S_S$ is in a single down-magnetized state and a $S_H$ down-magnetized domain is formed above. Just after switching off the field to zero, large magnetostatic fields act on the $S_S$ spin state just below the up-down domain boundary in $S_H$. As a consequence, a down-magnetized $S_H$ mirrored domain state is formed in $S_S$ and an up-magnetized decoration ring surrounds it (Fig. 8b). While the central down-down domain is self-stabilized, the decoration ring expands rapidly as long as the stray field remains large enough (Fig. 8c). Experiments show that the decoration ring reach rapidly a width of about 0.8 µm just after switching off the field (Fig. 6). This effect is essentially due to inter-stack dipolar stray fields that vanish rapidly in $S_S$ with the distance x from the border of the $S_H$ domain. The decoration ring is supposed to grow by propagation of the external wall, under the conjugated action of the static «orange peel» coupling, the magnetostatic stray field due to the finite size of the $S_H$ domain, and the intralayer magnetostatic interaction in $S_S$ [19], both favoring spin alignment in $S_S$ below $S_H$. More precisely, as soon as an up-magnetized state nucleates in an open channel of $S_S$, it propagates rapidly and tends to form a decoration ring around the black domain.

For a complementary approach with respect to our recent letter on the subject [19], and for a more quantitative approach, one considers the effect of the lateral variation of the resulting magnetostatic stray field, $H_{st} = H_{op} + H_z$ ($H_z$ being the sum of intra-stack and inter-stack magnetostatic fields), on a wall inside $S_S$. This wall stands at the lateral distance $x$ from the $S_H$ domain border for a channel-free down-down domain, or else from the base of a long channel (here with a typical length taken equal to 1 µm) (Fig. 9). Calculations have been performed for two superimposed 20 µm square down-magnetized domains in $S_S$ and $S_H$, or for non-reversed spin-up channels in $S_H$ with different widths $e$ (Fig. 9). They integrate effects due to each cobalt layer in the $S_H$ stack. $H_z$ can reach locally high values ($H_z$ = 800 Oe for x ~ 8 nm, i.e. close below the border of the $S_H$ domain (Fig. 10a). Large stray field values ($H_z$ > 400 Oe) act only over a restricted area in $S_S$, with width ~ 30 nm. This width has to be compared with the domain wall width estimated to $\Delta$ = 17 nm. Accounting that domain creation needs to overcome the wall bending energy, $H_z$ has to be large enough over extended regions (> 100 nm) for initiating a stable reversed up-magnetized domain in $S_S$. In agreement with this scheme, we experimentally found that nucleation takes place first inside channels, since the largest stray fields $H_z$ are acting there over the full channel length.

Since $H_z$ is highly negative just below the $S_H$ domain (Fig. 10a), mirroring spin-down domains are highly stabilized (Figs. 6, 8). Under the action of $H_{st}$, the decoration ring expands by propagation of its external wall in $S_S$ (Fig. 8b,c) at the expense of the down-up state. It slows down rapidly since the external wall is submitted to a stray field $H_z$ that decreases with x (Fig. 10b). Neglecting the «orange peel» coupling, the x- dependence of the perpendicular component is the sum of the non-uniform intra- and interlayer stray fields, $H_z$. It can be fitted with a good approximation by the $H_z \propto b / x$ law with b = 19 Oe.µm (Fig. 11). The resulting interaction field acting on the external wall thus decreases rapidly with the distance x as $H_{st}(x)$ = b/x. Moreover, the mean wall velocity in the presence of this weak resulting field is expected to slow down drastically when $H_{st}$ tends to zero, according to the expression:

$$v(x) = v_0 \exp[- A \, (H_{crit} / H_{st}(x))^{1/4}] \qquad (2)$$

Expression (2) is proposed by analogy to the "creep" expression (1) [17]. Assuming $H_{op} = 0$, the time "t" dependence of the width "x" of the decorated ring in zero field can be deduced:

$$t = 4/(v_0 \alpha^4) [6 - \exp(\alpha x^{1/4}) (6 - 6 \alpha x^{1/4} + 3 \alpha^2 x^{1/2} - \alpha^3 x^{3/4})] \qquad (3)$$

where $\alpha = A(H_{crit} / b)^{1/4}$. The calculated variation of x(t), i.e. the time dependence of the mean width of the decorated ring, just after switching off the field at t = 0, is plotted in Fig. 12. The experimental data points, deduced from image processing fit quite well the calculated x(t) curve when assuming $H_{op} \approx 0$. The high resolution of our CCD camera enables us to estimate the position of the wall in $S_S$, with an accuracy of 0.1 µm. The width x of the decorated ring reaches the value (0.8 ± 0.1) µm after t = 2 minutes (Fig. 12a), and afterwards evolves very slowly at long time (Fig. 12b). Since $H_{st}$ is never equal to zero, the wall continues to move over infinite times in the remanent state; in other words, it never reaches an equilibrium position.

**4. Domain decoration under magnetic field**

Starting from the domain pattern structure shown in Fig. 6, the width of the up-up magnetized decoration ring can be tuned by applying an additional field (Fig. 13). In these experiments, the field is maintained during the image acquisition (case (ii) described in section 2.2). Provided it displays the same positive sign as $H_{st}$, even a weak field (H = 7 Oe) helps significantly to propagate the external wall (Fig. 13a), giving an additional expansion of the decoration ring. In counterpart, a negative field tends to shrink the ring (Fig. 13b) since it competes with $H_{st}$. Since $H_{st}$ is strongly enhanced when the wall approaches the $S_H$ domain boundary, it becomes difficult to completely shrink the external ring wall. For high negative fields ($H < -100$ Oe), only channels maintain a white up-up spin configuration. In spite of a rather large applied negative field (H < - 200 Oe), these channels continue to maintain their up-up magnetized state in agreement with results presented in Fig. 10b. Dynamics can be well understood by replacing $H_{st}$ by $H_{st} + H$ in expression (2), and considering again the variation of $H_{st}$ with the wall position x. Typical experimental data on the time dependence of the width of the decoration ring in positive or negative applied field are reported in Fig. 14. At short time, the fast wall motion is due to efficient magnetostatic interactions (i.e. $H_{st}$). At long time, for $H > 0$, since $H_{st} \approx 0$, the quasi-linear dependence of x with t is essentially driven by the applied field (usual creep motion), and for $H < 0$, the wall is rapidly pushed down and stopped at the $S_H$ domain border.

**5. Proof of the strong mirroring effect: shrinking of the down-down domain state in positive field**

The goal of the study reported in this section is to check the strength of the magnetostatic coupling between the mirrored domains in $S_S$ and $S_H$, prepared as above (Fig. 6). For that purpose, we determined the evolution of this state with an applied positive magnetic field. After preparing a down-down (black) domain, a positive (22 Oe) magnetic field pulse was applied during 10 s. The obtained frozen remanent magnetic state was recorded (Fig. 15). This procedure was repeated for increasing field values up to 380 Oe, giving successive snapshots of the magnetic domain state (Fig 15). At low field, the central (black) domain is not modified, and as mentioned previously, one observes the expansion of the up-up magnetized decoration ring. $S_S$ is rapidly reversed in its up-spin state, and for $H >$ 45 Oe, no up-down state survives. This proves again the role of large local dipolar fields for stabilizing the down-down (black) domain state. Above 100 Oe, the up-up (white) channels broaden and new up-magnetized areas appear inside the down-down (black) state. As expected from the inspection of the $H_{st}$ profile (Fig. 10), when adding a large enough positive field contribution, the channel broadening and the nucleation of down-down internal domains are favored. This can be explained since either the magnetostatic stray field becomes partly compensated by the applied field at the center of the black domain or at wall boundaries. Nevertheless, the nucleation inside the black domain often needs an extra energy that is not necessarily brought by a small field. At large enough field, new channels can be formed inside the black domain from the above mentioned reversed areas. As a result, in a large field ($H \approx$ 167 Oe) (Fig. 15), the down-down state looks like a complex array of connected narrow ribbons; this structure is still stabilized by the remaining inter-stack local interactions. We could be tentatively tempted to compare this behavior with the field-induced domain phase diagram in highly anisotropic perpendicular films that is monitored by the intralayer magnetostatic interaction [25], but in our case, the most efficient effect comes from the interlayer magnetostatic coupling. The remaining magnetic (black) ribbons disappear progressively with increasing $H$, but in a non-uniform manner by successive jumps resulting of local depinning; during this process topological rearrangements are also involved. The mirroring effect between $S_S$ and $S_H$ domains or ribbons is so efficient that it acts up to the complete disappearance of the structure. As expected from the inspection of the $S_H$ hysteresis loop, the magnetization reversal is completed at 380 Oe to leave a saturated up-up domain state.

## 6. Magnetization reversal in high field

After saturating the bi-stack sample in a positive field and then slowly varying the applied negative field, the magnetization of $S_S$ becomes completely reversed down for $H \approx$ -100 Oe, i.e. well before to nucleate the first down-magnetized domains in $S_H$, that arises around - 220 Oe. We found that there is no correlation between nucleation sites in $S_S$ and $S_H$, nor with the position of the last vestigial 360° walls in $S_S$. Under $H \approx$ - 235 Oe, applied during 5 s, a few down-magnetized domains with jagged walls, develop in $S_H$ at the expense of the uniformly up-magnetized state. We found that down-down magnetized domains expand in the down-up state at a constant mean velocity, but only after a certain lag time $\tau$ that increases when reducing the field (for example, $\tau \approx$ 70 µs for $H =$ - 355 Oe). Thus, velocity measurements are not performed just after the switching of the field but as soon as the wall moves at constant velocity (t > $\tau$). Curiously, in spite of the presence of jagged walls, from the variation of the mean domain expansion velocity versus $H$, we can still define thermally activated and viscous regimes below and above $H_{crit}$ = 395 Oe.

Now, under a short high field ($H =$ - 458 Oe) pulse, we demonstrated experimentally that the magnetization reversal process differs markedly from that found above in the moderate field case. Remember that in moderate field, the magnetization is fully reversed down in $S_S$, before forming decorated mirrored reversed domains in $S_S$ and $S_H$ (Fig. 16a). Under a short high field pulse, the spin-up state in $S_S$, can be maintained a long time enough to allow a quite simultaneous nucleation of a few single down-magnetized reversed domains in $S_S$, (Fig. 16b), as well as mirrored down-down reversed domains in $S_S$ and $S_H$ (Fig. 16c). In othe words, these two states are then generated from an initially up-up single domain state (Fig. 17a). It is interesting to see that the two last types of domains can be nucleated at the same time at different places inside the film. Fortunately, even at such high field, the nucleation rate in the two $S_S$ and $S_H$ stacks is still rather low, so that one can investigate both the development of down-magnetized domains in $S_S$ (gray) keeping an $S_H$ up-magnetized (light gray) state (Figs. 16b and 17a), and of mirrored down-magnetized (black) domains in $S_S$ and $S_H$ (Fig. 16b and 17a). We found that the motion of the rough double wall separating the down-down and the up-up magnetized state (Fig. 16c) is faster than that found previously when starting from a down-up spin state (Fig. 16a). This is related to a decrease of the measured critical depinning field, $H_{crit}$ (210 Oe rather than 395 Oe); this phenomenon can be explained by the reduction of the inter-stack magnetostatic field at domain boundaries between spin configurations shown in Figs. 16a and 16c. Moreover, the mean velocity of walls separating the down-up (gray) and up-up (light gray), or the down-down (black) and up-up (light gray) states, has the same order of magnitude in a field of - 540 Oe; this gives rise to two different types of domains with close size (Fig. 16a,b). This is quite puzzling at first sight since the coercivity measured at low frequency differs markedly for the soft ($S_S$) and the hard ($S_H$) stacks. In spite of the lower depinning field for $S_S$, we demonstrated that this phenomenon can be explained by a higher domain wall velocity for coupled $S_H$ and $S_S$ stacks than only in $S_S$ when |$H$| becomes larger than 380 Oe. When the two types of domain approach to each other (Fig. 17b), one observes a slowing down of their wall motion. They don't merge at long time in zero field, preserving a narrow intermediate up-up magnetized region stabilized by the inter-stack magnetostatic coupling, in the same way as the decoration ring. This is consistent with our above scheme and calculations [19].

So, under a short high field pulse, nucleation of down-up (gray) or down-down (black) domain states can appear at the same time, and the reversal of a coupled $S_H$ and $S_S$ domain state by wall motion can be easier than that in $S_S$ alone. This does not contradict the fact that, at low field, the soft layer reverses its magnetization more easily than the hard layer.

## 7. Conclusions

We have reported on magnetic domain structures and dynamics in interacting ferromagnetic stacks (or layers) with perpendicular anisotropy. As for in-plane magnetized systems, interlayer magnetostatic stray fields can perturb the functioning of GMR or TMR devices. However, the mutual imprint of domains for in-plane magnetized films is ascribed to the stray field created by Néel walls [15]. In contrast, in films with out-of-plane magnetic anisotropy, not only Bloch walls play a role but this effect depends strongly of the finite domain size. Strong stray fields produced by a small down-magnetized domain in the hard stack favor the formation of a mirrored ferromagnetic domain in the soft layer, and even create an up-magnetized decoration ring in the soft layer at remanence. By considering both magnetostatic interactions and slow dynamics, we have shown how such decoration ring expands over infinite time, even in the absence of any applied field. More unexpected is that application of short large field pulses do not reverse instantaneously the magnetization in the soft layer but give rise to the simultaneous nucleation and propagation of reversed down-magnetized coupled domains in the soft and hard layers inside an up-up spin state. This emphasizes the notion that the hardness of a given layer in a bilayer soft-hard structure, has to be defined from a dynamic point of view. This has some implications for new generations of high speed devices. As devices continue to shrink in size, such local magnetostatic interactions and dynamic effects have to be considered in view of their implementation. Namely, the suppression of 360° domain walls in systems with out-of-plane anisotropy (induced by extrinsic defects, depending on growth or substrate cleaning procedures) might be sufficient to minimize the interlayer magnetostatic interaction. This is especially restrictive for the implementation of double coercivity based systems with in-plane anisotropy. More fast and ultrafast dynamic studies of interacting domain structures have to be carried out in thin layer structures, but also in associated patterned nano-size elements.

**Acknowledgements:** The authors wish to thank S. Auffret for assistance with film deposition, and A. Thiaville for enlightening discussions. This paper is a part of S. Wiebel's work done in the frame of a European fellowship supported by the ULTRASWITCH European project (HRPN-CT-2002-00318). This work has been performed in part during the early stage of the MAGLOG-STREP European program (FP-510993).

**Figure caption :**

Fig. 1 : Room temperature polar Kerr rotation (PKR) hysteresis loop ( — ) and minor loop (----) of the bi-stack structure. Field sweeping rate : 500 Oe/s.

Fig. 2 : Domain expansion in $S_S$: (a) initial domain state after switching off the field, (b) new snapshot after the application of a field pulse with magnitude $H = -74$ Oe and duration t = 6 ms, after switching off the field.

Fig. 3 : Variation of the mean domain wall velocity with $H$ for the $S_S$ stack in the bi-stack structure.

Fig. 4 : « Creep » plot of the mean domain wall velocity versus $H^{1/4}$ for the $S_S$ stack in the bi-stack structure.

Fig. 5 : Snapshots of the magnetic remanent domain state of $S_S$ after first saturating the sample in a positive field, applying a negative field $-H$, and switching off the field to zero. The following snapshots (a to d) were recorded after applying successively a field $H = -40.4$ Oe during 10 s (a), and 20s (b) ; $H = -45.7$ Oe during 10 s (c) ; $H = -51.1$ Oe during 2s. From the state (d), the field is reversed again to a positive value $H = 21.2$ Oe during 10 s (e). Image size : 30 μm x 30 μm.

Fig. 6 : PMOKE snapshot of the remanent magnetized state (observed 2 minutes after switching off the field to zero) imaged after applying a perpendicular field pulse $H = -235$ Oe during 5 s from a positive saturated single domain state. Image size 30μm x 30μm.

Fig. 7 : PMOKE image (30 μm x 30 μm) of the domain structure in the remanent state, observed one hour after the application of a pulse of field ($H = -272$ Oe) during 150 ms. Image size 30 μm x 30 μm

Fig.8: Sketch of the cross-section magnetic arrangement in the $S_H$ (top stack) and $S_S$ (bottom stack):

(a) : after the initiation of a $S_H$ domain,

(b) : expected remanent state, just after switching off the field,

(c) : expected remanent state after waiting a long time. The up-up ring is expanding through the down-up state. $H_C^S$ and $H_C^H$ are respectively the coercive field of $S_S$ and $S_H$.

Fig. 9 : Schematic upper view of a square down-down (black) magnetized domain with a non-reversed channel of width « e ». The distance x is taken either from the square side or the base of the channel.

Fig. 10 : Variation of the perpendicular component $H_z$ of the stray field with the distance x from the border of the $S_H$ domain or the base of channels with different width « e ». (a) at short scale, apart the limit of the $S_H$ domain for x = 0. (b) at longer scale, inside the channels.

Fig. 11 : Variation of the calculated stray field $H_z$ acting on the external wall in $S_S$ as a function of the distance x (——), and its best fit with the b / x law (-----).

Fig.12 : Calculated time dependence of the mean width x of the decorated ring in a linear (a) and a semi-logarithmic (b) plot. The position of the data points ▲ is compared to the calculated curve.

Fig. 13 : PMOKE snapshots obtained from the initial demagnetized domain state depicted in Fig. 6, (a ) : After applying a positive magnetic field $H$ = 7 Oe during 2 minutes, and freezing this state in zero field, (b) : In the presence of a negative field, $H$ = - 77 Oe, applied during 2 minutes. The image size is 30 μm x 30 μm.

Fig. 14 : Measured time dependence of the mean external wall displacement of the decorated up-up magnetized ring for two positive and one negative values of the perpendicularly applied field. The x value refers to the remnant ($H$ = 0) wall position measured at long time (t = 3 hours) Fig. 8). The lines are guides for the eyes.

Fig. 15: Schrinking of the down-down (black) domain structure in positive field, starting from an initial state obtained under a field pulse $H$ = - 194 Oe, applied during 12 μs. For all snapshots, the indicated field was applied during 10 s.
Image size : 30 μm x 30 μm for the two first snapshots, 20 μm x 20 μm for the others.

Fig. 16 : Cross view of the domain configuration when applying on a saturated up-up state:

(a) a relatively small negative field during a long time (idem as Fig. 8a),

(b) a larger negative field pulse with the formation of only a reversed domain in $S_S$,

(c) a larger negative field pulse with the formation of a mirrored down-down magnetized domain coupled state

Fig. 17: Snapshots of the remanent domain state after saturating first the sample in a positive field and applying a negative pulse of field $H = -540$ Oe with (a) 600 ns duration. (b) after an additionnal pulse of 600 ns duration. Note on (b) the stabilization of up-up (light grey) magnetized channels at the boundary between down-up and down-down domains. The nucleation of the down-up or down-down domains are located at their center. The radial orientation of the up-up magnetizated channels is clearly visible.
Image size: 66 μm x 42 μm.

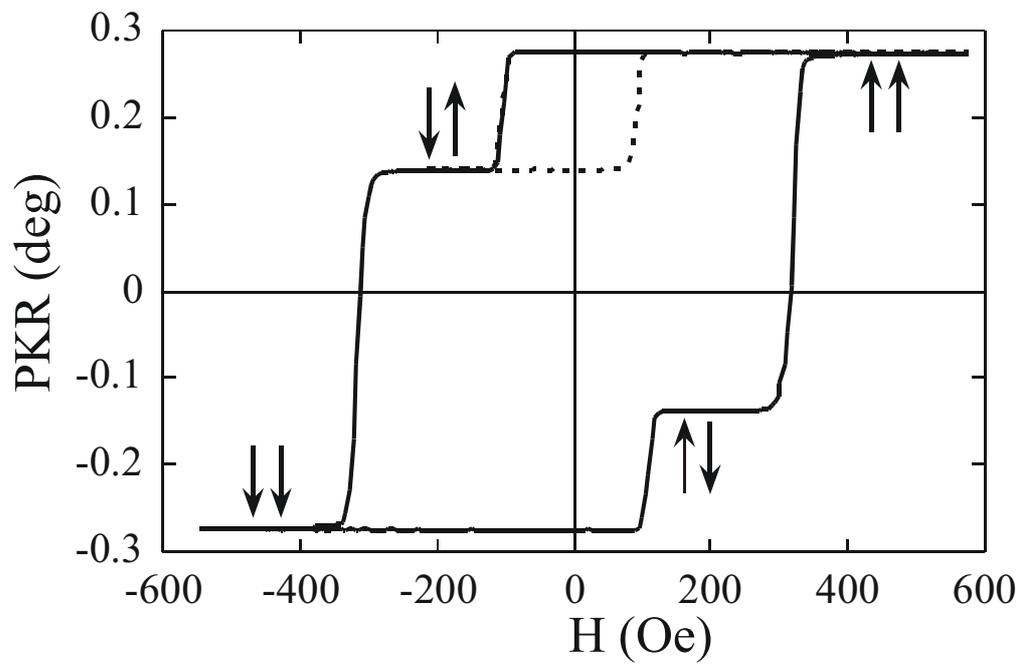

Fig. 1

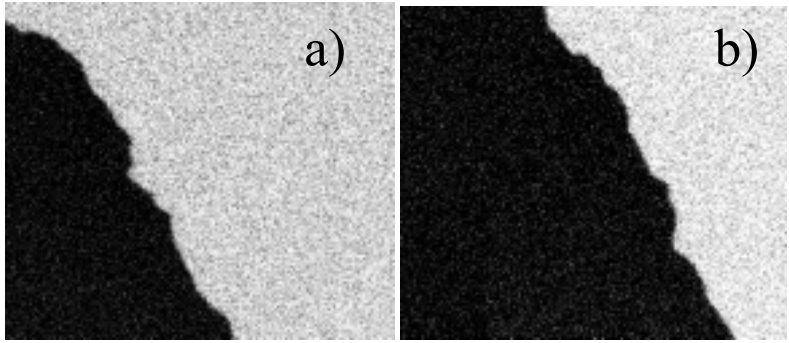

Fig.2

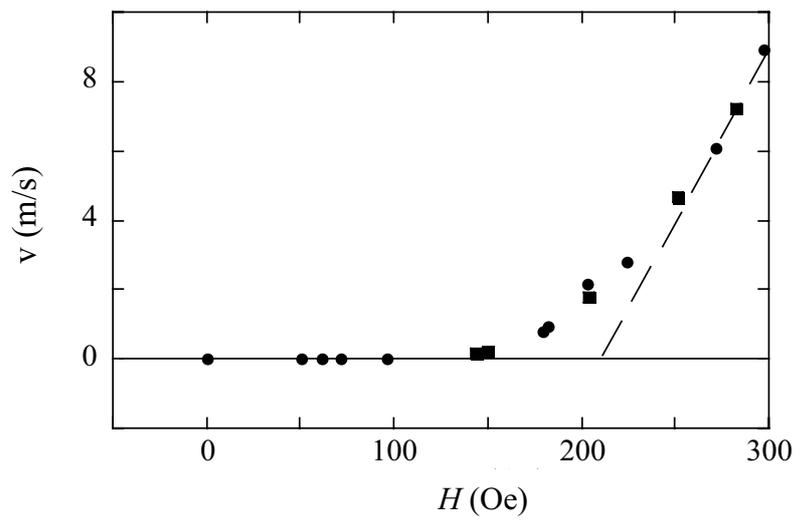

Fig. 3

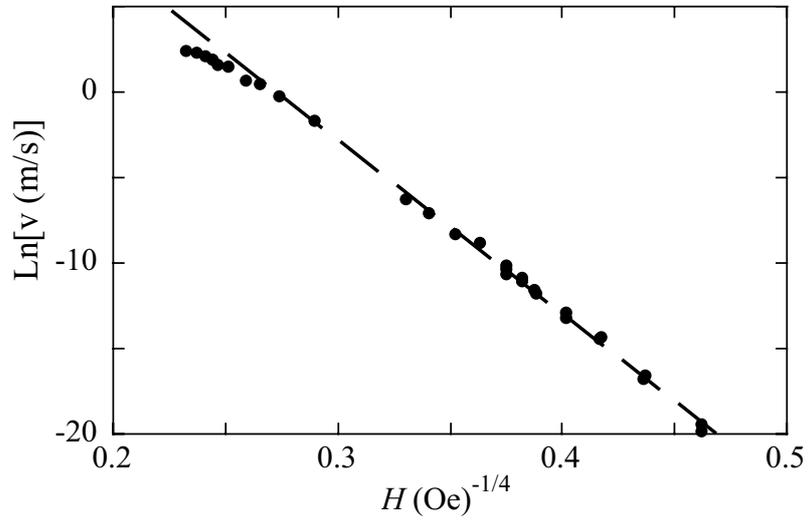

Fig. 4

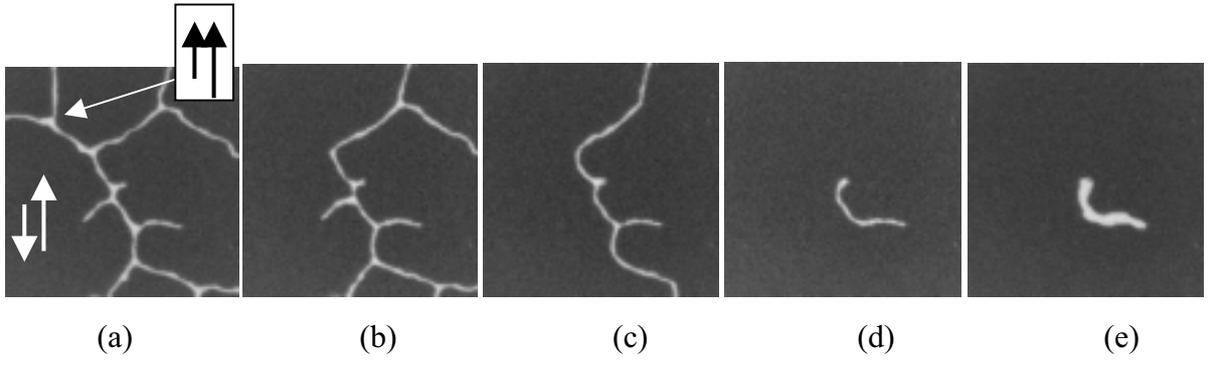

(a)　　　　　(b)　　　　　(c)　　　　　(d)　　　　　(e)

Fig. 5

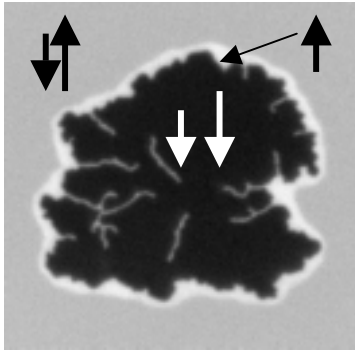

Fig. 6

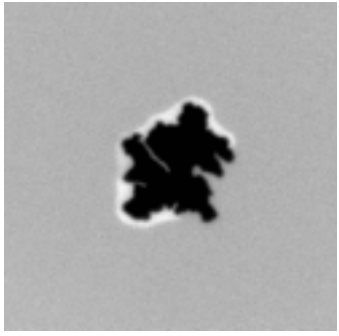

Fig. 7

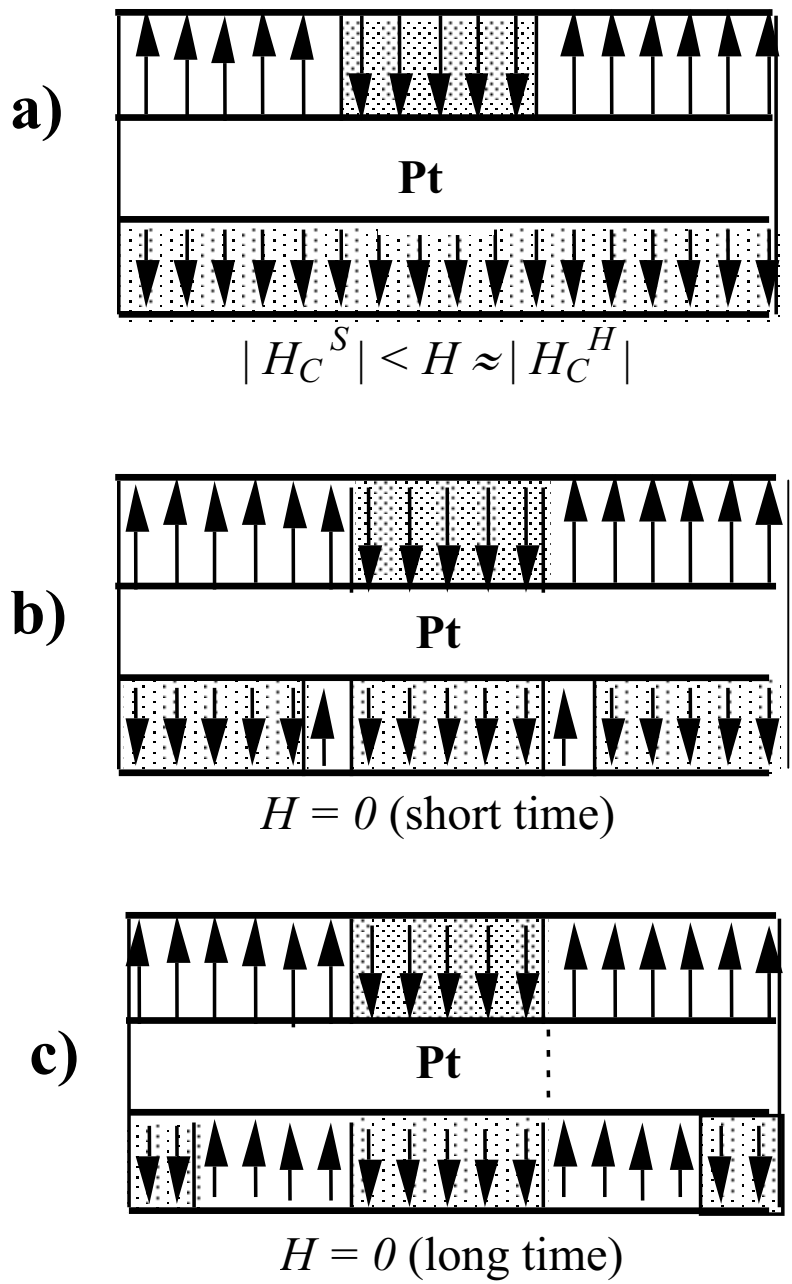

Fig. 8

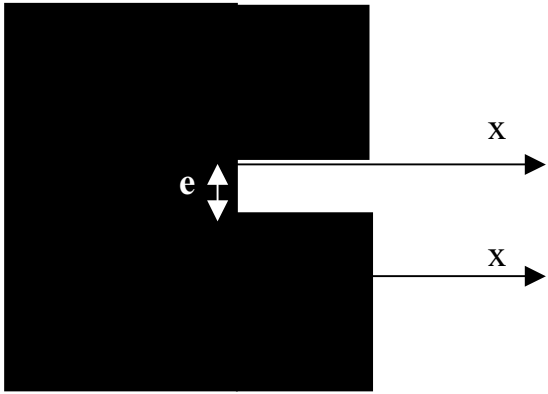

Fig. 9

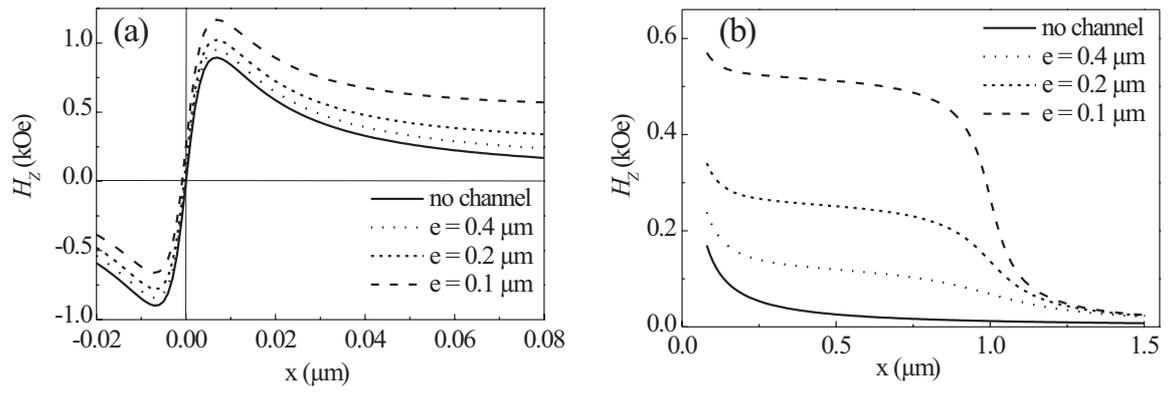

Fig. 10

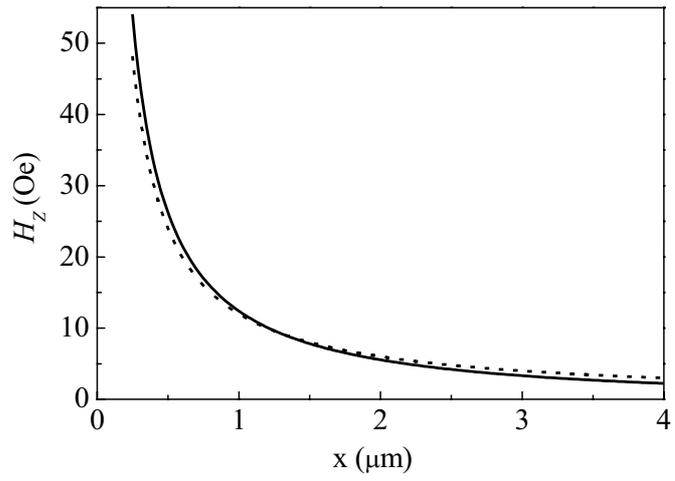

Fig. 11

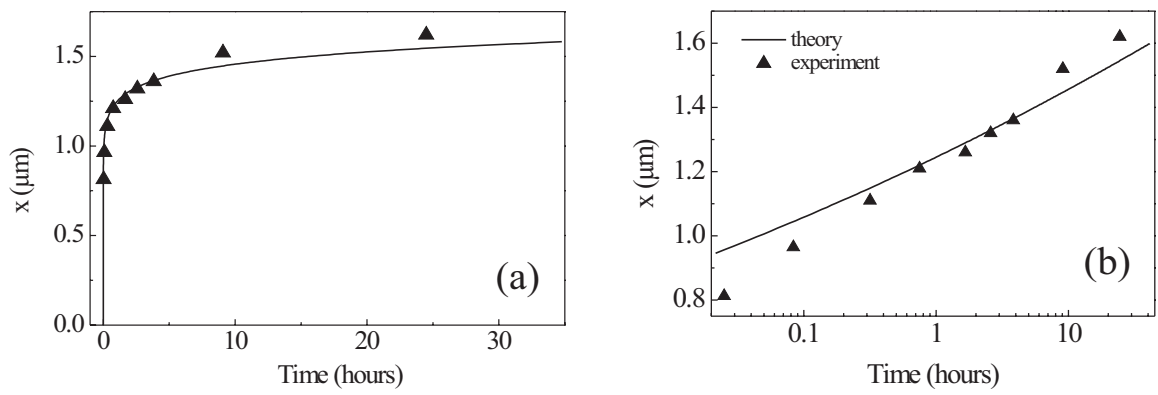

Fig. 12

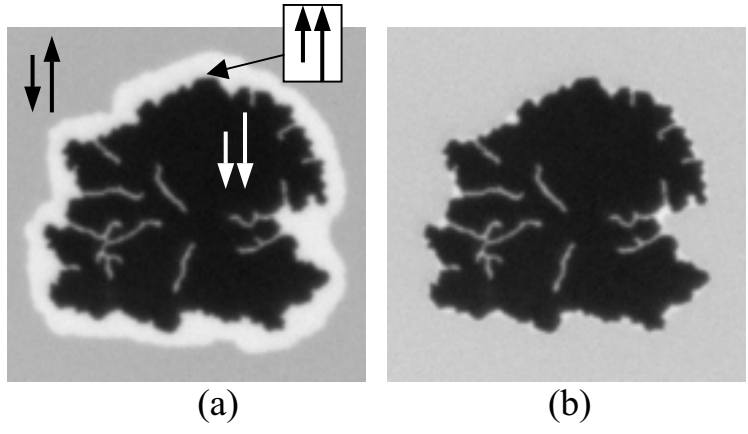

Fig. 13

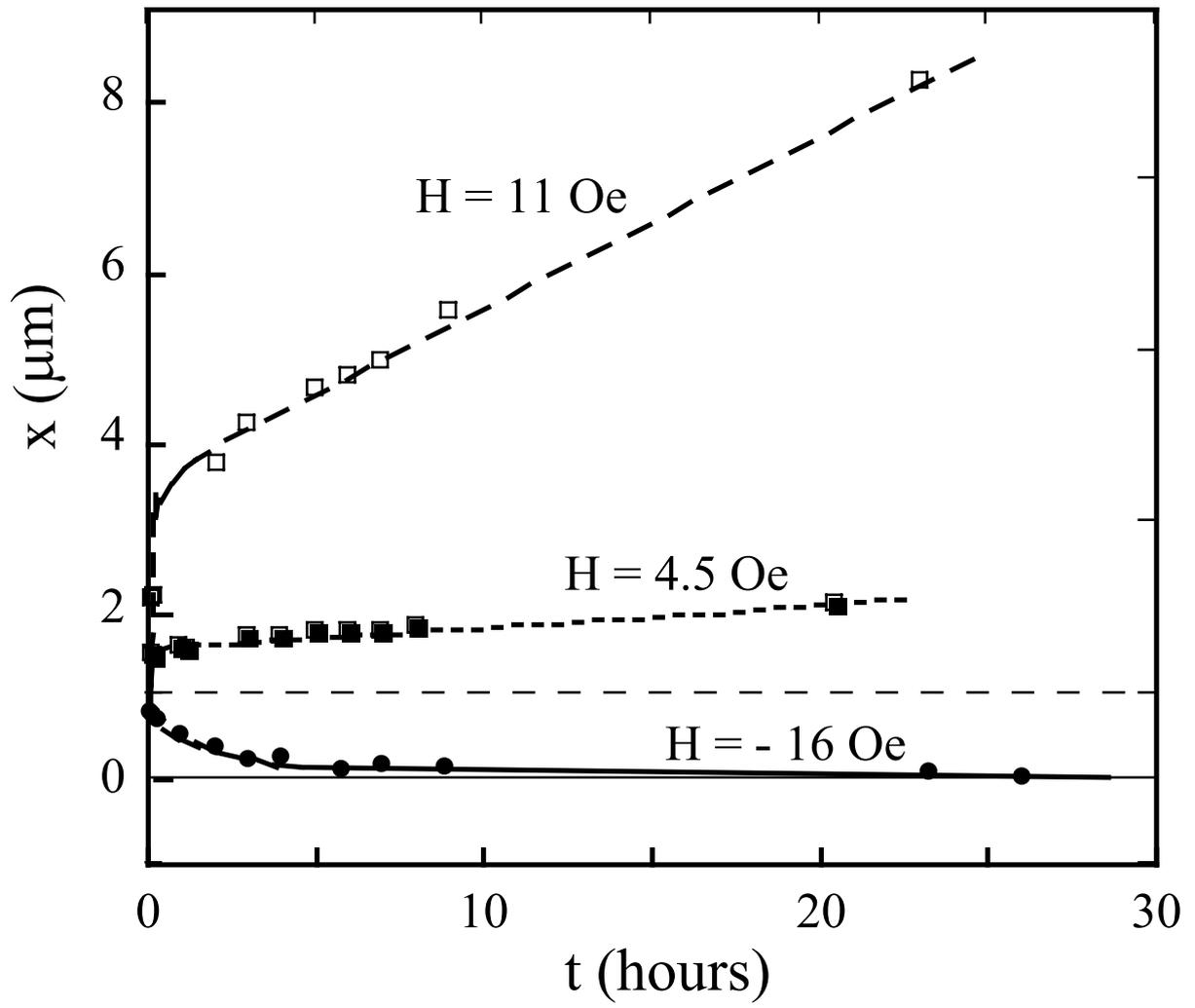

Fig. 14

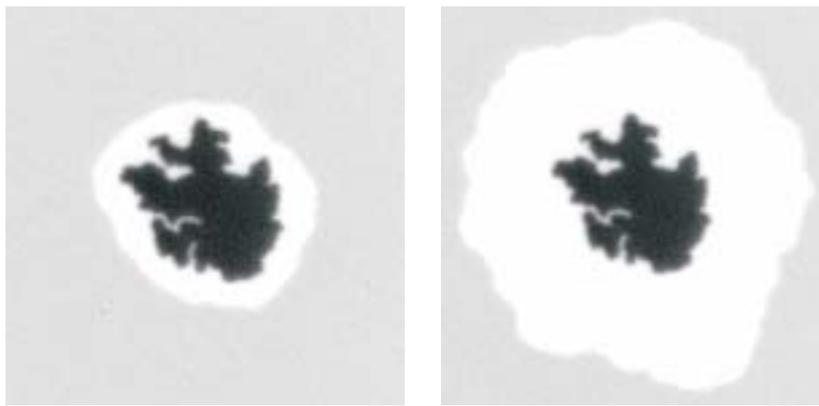

| H = 22 Oe | H = 33 Oe |

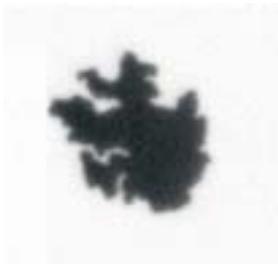 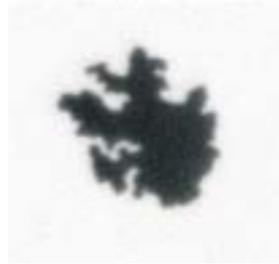 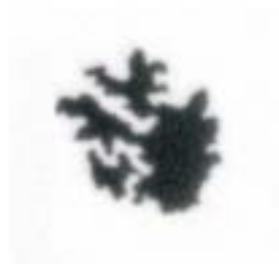 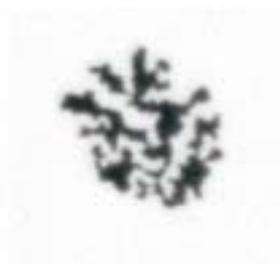

| H = 45 Oe | H = 100 Oe | H = 122 Oe | H = 145 Oe |

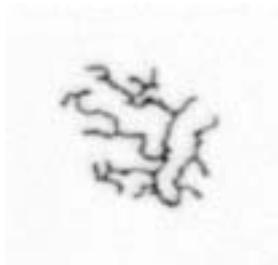 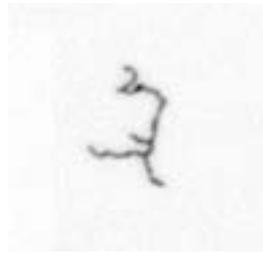 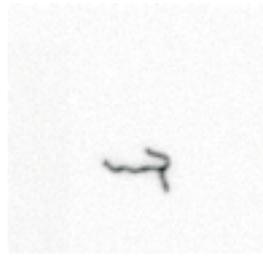 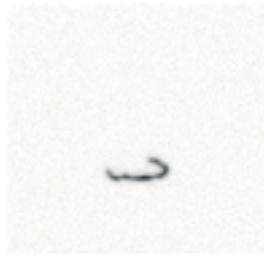

| H = 167 Oe | H = 300 Oe | H = 333 Oe | H = 366 Oe |

Fig. 15

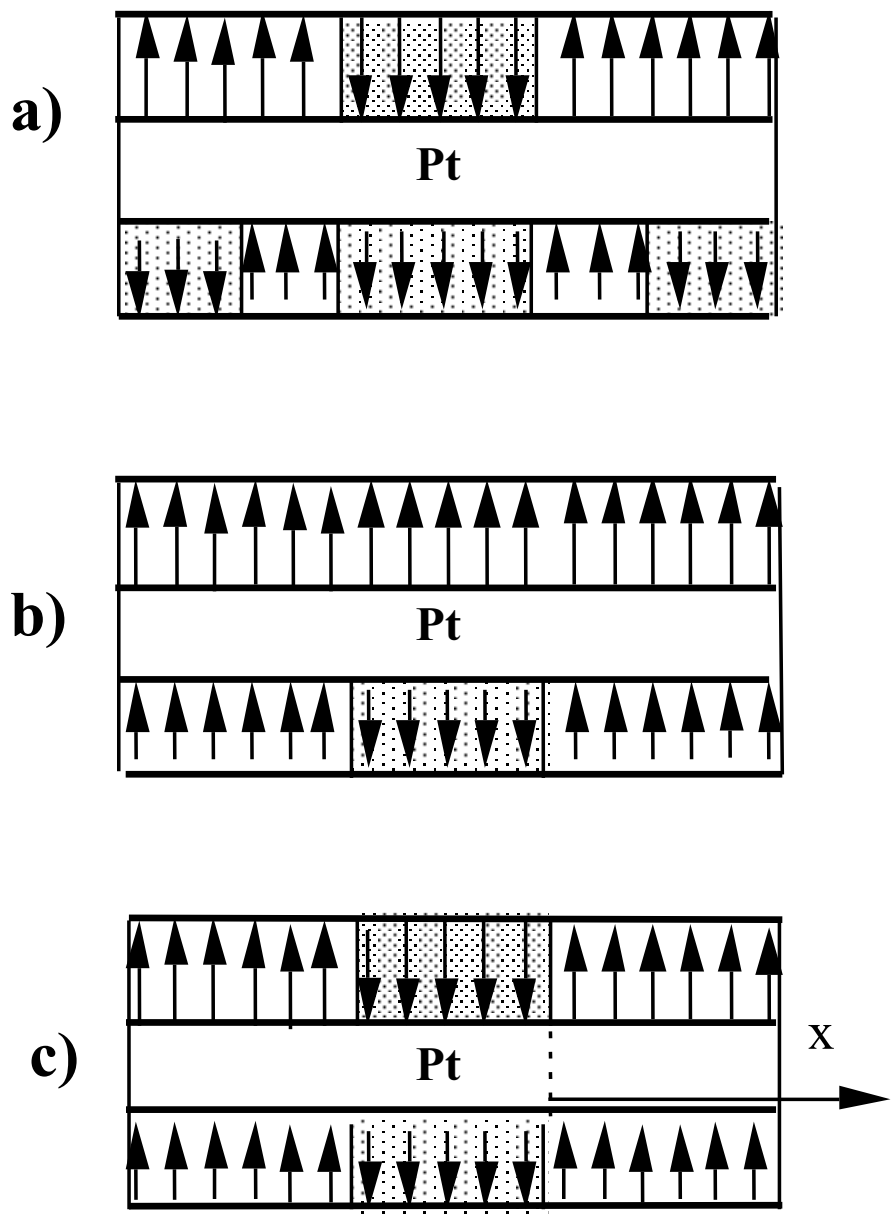

Fig. 16

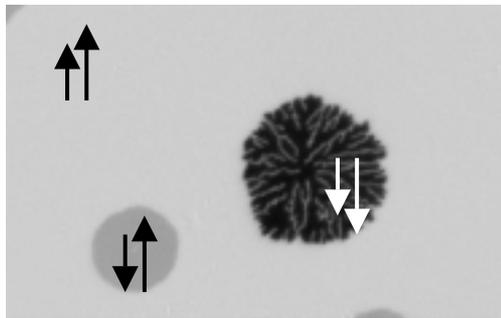 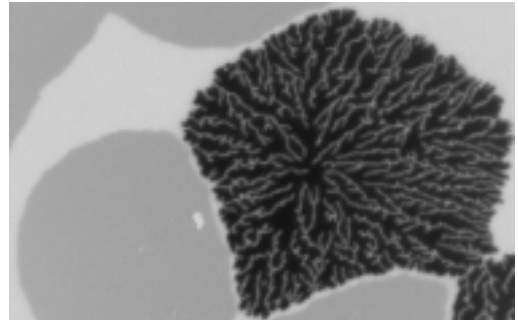

(a)             (b)

Fig. 17